\documentclass[12pt]{article} 
\begin{document} 

\begin{center}

{\bf \Large Impulse Gravity Generator Based on Charged  $YBa_2Cu_3O_{7-y}$ Superconductor with Composite Crystal Structure}

Evgeny Podkletnov$^1$, Giovanni Modanese$^2$

\bigskip

$^1$ {\it Moscow Chemical Scientific Research Centre \\
113452 Moscow - Russia} \\ E-mail: epodkletnov@hotmail.com

\bigskip

$^2$ {\it California Institute for Physics and Astrophysics \\
  366 Cambridge Ave., Palo Alto, CA 94306} \\ and \\
  {\it University of Bolzano -- Industrial Engineering \\
  Via Sernesi 1, 39100 Bolzano, Italy}
\\ E-mail: 
  giovanni.modanese@unibz.it

\bigskip

\end{center}

\begin{abstract} 
The detection of apparent anomalous forces in the vicinity of high-$T_c$ superconductors under non equilibrium conditions has stimulated an experimental research in which the operating parameters of the experiment have been pushed to values higher than those employed in  previous attempts. The results confirm  the existence of an unexpected physical interaction. An apparatus has been constructed and tested in which the superconductor is subjected to peak currents in excess of $10^4 \ A$, surface potentials in excess of 1 $MV$, trapped magnetic field up to 1 $T$, and temperature down to 40 $K$. In order to produce the required currents a high voltage discharge technique has been employed. Discharges originating from a superconducting ceramic electrode are accompanied by the emission of radiation which propagates in a focused beam without noticeable attenuation through different materials and exerts a short repulsive force on small movable objects along the propagation axis. Within the measurement error (5 to 7 \%) the impulse is proportional to the mass of the objects and independent on their composition. It therefore resembles a gravitational impulse. The observed phenomenon appears to be absolutely new and unprecedented in the literature. It cannot be understood in the framework of general relativity. A theory is proposed which combines a quantum gravity approach with anomalous vacuum fluctuations.
\end{abstract}

\setcounter{page}{1}
\tableofcontents

\section{Introduction}
\label{int}
 
Experiments showing possible anomalous forces between high-$T_c$ ceramic superconductors under non equilibrium conditions and test objects have been reported by several investigators since 1992 \cite{p1,p2,p3,p4,p5}. The observed phenomenology was difficult to explain and has been attributed to a so called ``gravity modification", because the reported effects mimic well the properties of the gravitational interaction, although their nature has never been clearly understood. In fact several alternative explanations of these results have been proposed \cite{p9,p10,p11,p12,Ummarino} in the attempt to bring the observed anomalies into the realm of known effects. 

Because of the great importance of any possible technical application of the reported effects, research activities have started in many laboratories since the first observation of the phenomenon \cite{p1}. Our recent research work focused on the improvement of the structure of the high-$T_c$ ceramic superconductors which have demonstrated capabilities of  creating anomalous forces. Moreover a  high-voltage discharge apparatus has been designed and constructed in order to easily reach those non equilibrium electromagnetic conditions that seem required to produce the force effects in  HTCs.

The results described in this report should be regarded as preliminary. An improved version of the experiment is currently being planned. Nevertheless, the body of results, as well as the complexity of the experimental procedures and of the theoretical interpretation are such that a detailed description and diffusion could not be further delayed. All measurements were done by E. Podkletnov in Moscow, while G. Modanese provided theoretical advice.

\section{Experimental}
\label{exp}

\subsection{General description of the installation}

The initial variant of the experimental set-up was based on a high-voltage generator placed in a closed cylinder chamber with a controlled gas atmosphere, as shown in Fig.\ 1. Two metal spheres inside the chamber were supported by hollow ceramic insulators and had electrical connections that allowed to organize a discharge between them, with voltage up to 500 $kV$. One of the spheres had a thin superconducting coating of $YBa_2Cu_3O_{7-y}$ obtained by plasma spraying using a ``Plasmatech 3000S" installation. This sphere could be charged to high voltage using a high voltage generator similar to that of Van de Graaf. The second sphere could be moved along the axis of the chamber, the distance between the spheres varying from 250 to 2000 $mm$. Spheres with a diameter from 250 to 500 $mm$ were used in the experiment. It was possible to fill the chamber with helium vapours or to create rough vacuum using a rotary pump. The walls of the chamber were made of non-conducting plastic composite material, with a big quartz glass window along one of the walls which allowed to observe the shape, the trajectory and the colour of the discharge. In order to protect the environment and the computer network from static electricity and powerful electromagnetic pulses, the chamber could be shielded by a Faraday cage with cell dimensions of $2.0 \times 2.0~cm$ and a rubber-plastic film material absorbing ultra high frequency (UHF) radiation.

The superconducting sphere was kept at a temperature between 40 and 80 $K$, which was achieved by injecting liquid helium or liquid nitrogen through a quartz tube inside the volume of the superconducting sphere before the charging began. The inside volume of the chamber was evacuated or filled with helium in order to avoid the condensation of moisture and different gases on the superconducting sphere. The temperature of the superconductor was measured using a standard thermocouple for low temperature measurements and was typically around 55-60 $K$. Given the good heat conductivity of the superconductor, we estimated that the temperature difference in the ceramic did not exceed 1 $K$. 

An improved variant of the discharge chamber is shown in Fig.\ 2. The charged electrode was changed to a toroid attached to a metal plate and a superconducting emitter which had the shape of a disk with round corners. The non-superconducting part of the emitter was fixed to a metal plate using metal Indium or Wood's metal, the superconducting part of the emitter faced the opposite electrode. The second electrode was a metal toroid of smaller diameter, connected to a target. The target was a metal disk with the diameter of 100 $mm$ and the height of 15 $mm$. The target was attached to a metal plate welded to the toroid.

This improved design of the generator was able to create a well-formed discharge between the emitter and the target, still the trajectory was not always repeatable and it was difficult to maintain constant values of current and voltage. The chamber was also not rigid enough to obtain high vacuum and some moisture was condensing on the emitter, damaging the superconducting material and affecting the discharge characteristics. The large distance between the electrodes also caused considerable dissipation of energy during discharge. In order to improve the efficiency of operation, the measuring system and the reproducibility of the discharge, an entirely new design of the vacuum chamber and the charging system was created. 

The final variant of the discharge chamber is presented in Fig.\ 3 (the apparatus is shown in a vertical position though actually it is situated parallel to the floor). This set-up allowed to reduce the dimensions of the installation and to increase the efficiency of the process. The chamber has the form of a cylinder with the approximate diameter of 1 $m$ and the length of 1.5 $m$ and is made of quartz glass. The chamber has two connecting sections with flanges which allow to change the emitter easily. The design permits to create high vacuum inside or to fill the whole volume with any gas that is required. The distance of the discharge has been decreased considerably giving the possibility to reduce energy dissipation and to organize the discharge in a better way. The distance between the electrodes can vary from 0.15 to 0.40 $m$ in order to find the optimum length for each type of the emitter. 

The discharge can be concentrated on a smaller target area using a big 
solenoid with the diameter of 1.05 $m$ that is wound around the chamber using copper wire with the diameter of 0.5 $cm$. The magnetic flux density is 0.9 $T$. A small solenoid is also wound around the emitter (Fig.\ 3) so that the magnetic field can be frozen inside a superconductor when it is cooled down below the critical temperature.

The refrigeration system for the superconducting emitter provides a sufficient amount of liquid nitrogen or liquid helium for the long-term operation and the losses of gas due to evaporation are minimized because of the high vacuum inside the chamber and thus of a better thermal insulation.

A photodiode is placed on the transparent wall of the chamber and is connected to an oscilloscope, in order to provide information on the light parameters of the discharge. Given the low pressure and the high applied voltage, emission of X-rays from the metallic electrode cannot be excluded, but the short duration of the discharge makes their detection difficult. Use of a Geiger counter and of X-rays sensitive photographic plates did not yield any clear signature of X-rays.

A precise measurement of the voltage of the discharge is achieved using a capacitive sensor that is connected to an oscilloscope with a memory option as shown in the upper part of Fig.\ 3. Electrical current measurements are carried out using a Rogowski belt, which is a single loop of a coaxial cable placed around the target electrode and connected to the oscilloscope. 

The old fashioned Van de Graaf generator used in the previous stage of this work was replaced by a high voltage pulse generator as shown in Fig.\ 4. This pulse generator is executed according to the scheme of Arkadjev-Marx and consists of twenty capacitors (25 $nF$ each) connected in parallel and charged to a voltage up to 50-100 $kV$ using a high voltage transformer and a diode bridge. The capacitors are separated by resistive elements of about 100 $k\Omega$. The scheme allows to charge the capacitors up to the needed voltage and then to change the connection from a parallel to a serial one. The required voltage is achieved by changing the length of the air gap between the contact spheres C and D. A syncro pulse is then sent to the contacts C and D which causes an overall discharge and serial connection of the capacitors and provides a powerful impulse up to 2 $MV$ which is sent to the discharge chamber. The use of such an impulse generator allows for a precisely controlled voltage, much shorter charging time and good reproducibility of the process.

\subsection{Superconducting emitter, fabrication methods}

The superconducting emitter has the shape of a disk with the diameter of 80-120 $mm$ and the thickness of 7-15 $mm$. This disk consists of two layers: a superconducting layer with chemical composition $YBa_2Cu_3O_{7-y}$ (containing small amounts of $Ce$ and $Ag$) and a normal conducting layer with chemical composition $Y_{1-x}Re_xBa_2Cu_3O_{7-y}$, where $Re$ represents $Ce$, $Pr$, $Sm$, $Pm$, $Tb$ or other rare earth elements. The materials of both layers were synthesized using a solid state reaction under low oxygen pressure (stage 1), then the powder was subjected to a melt texture growth (MTG) procedure (stage 2). Dense material after MTG was crushed, ground and put through sieves in order to separate the particles with the required size. A bi-layered disk was prepared by powder compaction in a stainless steel die and sintering using seeded oxygen controlled melt texture growth (OCMTG) (stage 3). For the emitters with the diameter of 120 $mm$ usual sintering was applied instead of seeded OCMTG (stage 4). After mechanical treatment  the ceramic emitter was attached to the surface of the cooling tank in the discharge chamber using Indium based alloy. 

{\bf Stage 1 - }   Micron-size powders of $Y_2O_3$ and $CuO$, $BaCO_3$ were mixed in alcohol for 2 hours, then dried and put in zirconia boats in a tube furnace for heat treatment. The mixture of powders was heated to 830 $^o${\it C} and kept at this temperature for 8 hours at oxygen partial pressure of $2.7 \cdot 10^2~Pa$ (or 2-4 $mBar$) according to \cite{p13,p14}. The material of the normal conducting layer was sintered in a similar way.

{\bf Stage 2 - }  Micron-size powder of $YBa_2Cu_3O_x$ was pressed into pellets using a metal die and low pressure. The pellets were heated in air to 1050 $^o${\it C} (100 $^o${\it C} per hour) then cooled to 1010 $^o${\it C} (10 $^o${\it C} per hour) then cooled to 960 $^o${\it C} (2 $^o${\it C} per hour) then cooled to room temperature (100 $^o${\it C} per hour) according to a standard MTG technique \cite{p15,p16}. The quantity of 211 phase during heating was considerably reduced and the temperature was changed correspondingly. $ReBa_2Cu_3O_{7-x}$ was also prepared using MTG, but the temperature was slightly changed according to the properties of the corresponding rare earth oxide.

{\bf Stage 3 - }  Bulk material after MTG processing was crushed and ground in a ball mill. The particles with the size less than 30 $\mu m$ were used for both layers of the ceramic disk. The particles were mixed with polyvinyl alcohol binder. The material of the first layer was put into a die, flattened and then the material of the second layer was placed over it. The disk was formed using a pressure of 50 $MPa$. The single crystal seeds of $Sm123$ (about 1 $mm^3$) were placed on the surface of the bi-layered disk so that the distance between them was about 15 $mm$ and the disk was subjected to a OCMTG treatment in 1\% oxygen atmosphere. The growth kinetics of YBCO superconductor were controlled during isothermal melt texturing. A modified melt texturing process was applied, where instead of slow cooling following melting, isothermal hold was employed in the temperature range where the growth is isotropic. By this modification, the time required to texture the disk was reduced to 7 hours which is about 10 times faster than a typical slow cooling melt texturing process.  The crystallization depth was controlled by applying the corresponding temperature and time parameters. Cubic $Sm123$ seeds were obtained using the nucleation and growth procedure as described in \cite{p19,p20}. A thin layer of the material was removed from the top surface of the disk to a depth of 0.3 $mm$ and the edges of the upper surface were rounded using diamond tools.

{\bf Stage 4 - }  For the emitters with the diameter of 120 $mm$ it is technically difficult to apply seeded MTG method, therefore normal sintering was carried out. Bulk material after MTG processing (after stage 2) was crushed and sieved and the following size particles were used for both layers of the ceramic disk, the amount is given in weight percent:  

\begin{tabular}{cc}
500-400 $\mu m$ & \ \ \ 50-60\% \\
120-60  $\mu m$ & \ \ \ 25-35\% \\
$< 20$ $\mu m$ & \ \ \ 15-25\% \\
\end{tabular}

The particles were mixed according to the ratio listed above using polyvinyl alcohol as a binder. The material of the first layer was put into a die, flattened and the material of the second layer was placed over it. The disk was formed using the pressure of 120 $MPa$ and sintered in oxygen at 930 $^o${\it C} for 12 hours followed by slow cooling down to room temperature. The edges of the upper surface were rounded using diamond tools. 

X-ray diffraction, transition temperature, electrical conductivity and critical current density were measured for both layers of various emitters using standard techniques. 

\subsection{Organization of the discharge and measurements of the effect.}

The discharge chamber is evacuated to 1.0 $Pa$ using first a rotary pump and then a cryogenic pump. When this level of vacuum is reached, liquid nitrogen is pumped into a tank inside the chamber that contacts the superconducting emitter. Simultaneously a current is sent to the solenoid that is wound around the emitter, in order to create a magnetic flux inside the superconducting ceramic disk. When the temperature of the disk falls below the transition temperature (usually 90 $K$) the solenoid is switched off. The experiment can be carried out at liquid nitrogen temperatures or at liquid helium temperatures. If low temperatures are required, the tank is filled with liquid helium and in that case the temperature of the emitter reaches 40-50 $K$.

The high voltage pulse generator is switched on and the capacitors are charged to the required voltage. It takes about 120 $s$ to charge the capacitors. A syncro pulse is sent to a pair of small metal spheres marked as C and D in Fig.\ 5. A discharge with voltage up to 2 $MV$ occurs between the emitter and the target. Half a second before the discharge, a short pulse of direct current is sent for 1 $s$ to the big solenoid that is wound around the chamber, in order to concentrate the discharge and to direct it to the same area on the target electrode. This pulse lasts for only 1 $s$ not to cause the overheating of the big solenoid.

The effects are measured along the projection of the axis line which connects the center of the emitter with the center of the target. Laser pointers were used to define the projection of the axis line and impulse sensitive devices were situated at the distance of 6 $m$ and 150 $m$ from the installation (in another building across the area). 

Normal pendulums were used to measure the pulses of gravity radiation coming from the emitter. The pendulums consisted of  spheres of different materials hanging on cotton strings inside glass cylinders under vacuum. One end of the string was fixed to the upper cap of the cylinder, the other one was connected to a sphere. The spheres had typically a diameter from 10 to 25 $mm$ and had a small pointer in the bottom part. A ruler was placed in the bottom part of the cylinder, 2 $mm$ lower than the pointer. The deflection was observed visually using a ruler inside the cylinder (Fig.\ 5). The length of the string was typically 800 $mm$, though we also used a string 500 $mm$ long. Various materials were used as spheres in the pendulum: metal, glass, ceramics, wood, rubber, plastic. The tests were carried out when the installation was covered with a Faraday cage and UHF radiation absorbing material and also without them. The installation was separated from the impulse measuring devices situated 6 $m$ away by a brick wall of 0.3 $m$ thickness and a list of steel with the dimensions $1~m \times 1.2~m \times 0.025~m$. The measuring systems that were situated 150 $m$ away were additionally shielded by a brick wall of 0.8 $m$ thickness.

In order to define some other characteristics of the gravity impulse - in particular its frequency spectrum - a condenser microphone was placed along the impact line just after the glass cylinders. The microphone was connected to a computer and placed in a plastic spherical box filled with porous rubber. The microphone was first oriented with a membrane facing the direction of the discharge, then it was turned 22.5 degrees to the left, then 45 degrees to the left, then 67.5 degrees and finally 90 degrees. Several discharges were recorded in all these positions at equal discharge voltage.

\section{Results}
\label{res}

Several unexpected phenomena were observed during the experiments. The discharge in the installation corresponding to the initial set-up (Fig. 1) at room temperature in the voltage range from 100 $kV$ to 450 $kV$ was similar to a discharge with non-coated metal spheres and consisted of a single spark between the closest points on the spheres. When the superconductor coated sphere was cooled down below the transition temperature, the shape of the discharge changed in such a way that it did not form a direct spark between two spheres, but the sparks appeared from many points on the superconducting sphere and then moved to the corresponding electrode. When the voltage was over 500 $kV$ the discharge at the initial stage had a tendency to cause some glow with the shape of a hemisphere. This glow separated from the sphere and then broke into multiple sparks which combined into more narrow bundle and finally hit the surface of the target electrode. 

Repeated discharges at high voltages caused damage to the superconducting coating and partial separation of the ceramic material from the metal sphere, as the refrigeration system was not efficient enough. Also the direction of the discharge was not always repeatable. The experiments were continued with the improved variant of the installation as shown in Fig.\ 2 and then with the final variant of the installation as shown in Fig.\ 3. This new configuration allowed to increase the reproducibility of the discharge and the superconducting emitter was not damaged with high voltage. With voltage lower than 400 $kV$ the discharge had the shape of  a spark but when the voltage was increased to 500 $kV$ the front of the moving discharge became flat with diameter corresponding to that of the emitter. This flat glowing discharge separated from the emitter and moved to the target electrode with great speed. The whole time of the discharge as defined by the photo diode was between $10^{-5}$ and $10^{-4}~s$. The peak value of the current at the discharge for the maximum voltage ($2 \cdot 10^6 \ V$) is of the order of $10^4 \ A$.

It was found that high voltages discharges organized through the superconducting emitter kept at the temperature of 50-70 $K$ were accompanied by a very short pulse of radiation coming from the superconductor and propagating along the axis line connecting the center of the emitter and the center of the target electrode in the same direction as the discharge. The radiation appeared to penetrate through different bodies without any noticeable loss of energy. It acted on small interposed mobile objects like a repulsive force field, with a force proportional to the mass of the objects. As the properties of this radiation are similar to the properties of the gravity force, the observed phenomenon was called a gravity impulse.

In order to investigate the interaction of this gravity impulse with various materials, several tests were carried out, with pendulums and microphones, as described in the experimental part. The deflection of the pendulum was observed visually (see Fig.\ 5) and the corresponding $\Delta l$ value was measured as a function of the discharge voltage. The correlation between the discharge voltage and the corresponding deflection of the pendulum as measured for two different emitters is listed in Tables 1, 2. Each value of $\Delta l$ that is given in the table represents the average figure calculated from 12 discharges. A rubber sphere with a weight of 18.5 grams was used as material of the pendulum for the data listed in Tables 1, 2. The deflection caused an alteration in the potential energy of the pendulum which was proportional to $\Delta h$ as shown in the table. A graphic illustration of this dependence for two different emitters is given in Fig.\ 6.

\bigskip

\begin{table}
\caption{Emitter N.\ 2. Influence of high voltage discharges on the deflection of the pendulum. Experimental data are the average of 12 measurements. The standard deviation of the single data is between 5 and 7 \%.
\label{t1}}
\begin{tabular}{cccc}
& & & \\
Voltage ($kV$) & $\Delta l~(mm)$ & $\Delta h~(mm)$ & Estimated $\Delta E~(J \cdot 10^{-4})$ \\
500 & 56.5 & 2.0 & 3.6 \\
750 & 91.3 & 5.2 & 9.5 \\
1000 & 110.4 & 7.7 & 13.9 \\
1250 & 123.0 & 9.5 & 17.3 \\
1500 & 131.6 & 10.9 & 19.8 \\
1750 & 137.6 & 11.9 & 21.7 \\
2000 & 142.0 & 12.7 & 23.1 \\
\end{tabular}
\end{table}

\bigskip

\begin{table}
\caption{Same for Emitter N.\ 1.
\label{t2}}
\begin{tabular}{cccc}
& & & \\
Voltage ($kV$) & $\Delta l~(mm)$ & $\Delta h~(mm)$ & Estimated $\Delta E~(J \cdot 10^{-4})$ \\
500 & 40.0 & 1.0 & 1.8 \\
750 & 70.9 & 3.2 & 5.7 \\
1000 & 85.3 & 4.6 & 8.3 \\
1250 & 94.6 & 5.6 & 10.2 \\
1500 & 100.8 & 6.4 & 11.6 \\
1750 & 104.7 & 6.9 & 12.5 \\
2000 & 107.1 & 7.2 & 13.1 \\
\end{tabular}
\end{table}

\bigskip

Both emitters, N.\ 1 and N.\ 2, were manufactured using the same OCMTG technology, but the thickness of the superconducting layer was equal to 4 $mm$ for the emitter N.\ 1 and 8 $mm$ for the emitter N.\ 2. Emitter N.\ 2 could be magnetized to a much higher value. The thickness of the normal conducting layer has a smaller influence on the force of the gravity impulse, but for better results the thickness should be bigger than 5 $mm$.

It was found that the force of the impact on pendulums made of different materials does not depend on the material but is only proportional to the mass of the sample. Pendulums of different mass demonstrated equal deflection at constant voltage. This was proved by a large number of measurements using spherical samples of different mass and diameter. The range of the employed test masses was between 10 and 50 grams. It was also found that there exist certain deviations in the force of the gravity impulse within the area of the projection of the emitter. These deviations (up to 12-15\% max) were found to be connected with the inhomogenities of the emitter material and various imperfections of the crystals of the ceramic superconductor, and with the thickness of the interface between superconducting and normal conducting layers. 

Measurements of the impulse taken at close distance (3-6 $m$) from the installation and at the distance of 150 $m$ gave identical results, within the experimental errors. As these two points of measurements were separated by a thick brick wall and by air, it is possible to admit that the gravity impulse was not absorbed by the media, or the losses were negligible. 

The force ``beam" obtained with the latest experimental set-up does not appear to diverge and its borders are clear-cut. However, considerable efforts were necessary in order to concentrate the radiation and reach a good reproducibility. As mentioned above, the direction emission always coincides with the direction of the discharge. In the initial experiments (with the Van den Graaf generator), the direction of the beam varied, depending on the direction of the discharge, as the sparks moved to different points on the superconducting sphere. Later it was found that the magnetic field created by the solenoid wound around the chamber is able to concentrate the discharge and to direct it to the same area on the superconducting electrode.

The bi-layered emitters used in this experiment were mainly of two types. The first one was obtained after Stage 3 and had a structure typical for multiple-domain levitators with well crystallized and oriented grains of the superconducting layer. The second type was made by the material obtained after Stage 4 and consisted of densely packed non oriented polycrystalline structure in both layers. The superconducting layer in both types of emitters consisted of $YBa_2Cu_3O_{7-y}$ orthorhombic superconductor with lattice parameters $a=3.89 \AA$, $b=11.69 \AA$, $c=3.82 \AA$. The addition of small amounts of $CeO_2$ led to an improvement in the magnetic flux pinning properties of the Y123 compound. The emitter obtained by seeded MTG process had a superconducting layer with a maximum trapped field of 0.5 $T$ at 77 $K$ and a critical current density in excess of $5 \cdot 10^4~A/cm^2$. The transition temperature varied from 87 to 90 $K$ with a transition width of about 2 degrees. The normal conducting layer had crystal lattice parameters close to those of the superconductor: $a=3.88 \AA$, $b=11.79 \AA$, $c=3.82 \AA$. Both layers demonstrated high electrical conductivity (over 1.5 $Sm$) at room temperature and the $Y_{1-x}Re_xBa_2Cu_3O_{7-y}$ layer was a normal conductor above 20 $K$.

In general the emitters obtained by seeded OCMTG  were much more efficient than the sintered bi-layered emitters and allowed to obtain a much stronger radiation impulse, contained by the projection of the emitter. The sintered bi-layered emitters had much lower values of the trapped magnetic field and this yielded bigger energy dissipation and a weaker gravity impulse. The only advantage of the sintered emitters is that a weak impulse effect is always present, probably because the diffusion between the layers is limited. The emitters with oriented crystal structure (after seeded MTG) can be easily spoiled if the interface thickness between the two layers reaches a certain value, therefore the temperature and time parameters should be monitored carefully in order to limit the interaction between the two layers. It was found that the seeded crystal growth should be stopped before it reaches the interface region. 

It was also found that the gravity impulse was to some extent proportional to the magnetic field inside the superconductor, which was created using a small solenoid during cooling down to liquid nitrogen temperature. Therefore, at recent stages of the experiment the solenoid was replaced by a powerful permanent NdFeB magnet (50 $MOe$) with a diameter corresponding to the diameter of the emitter and a thickness of 20 mm. This disk-shaped magnet was attached with one surface to the cooling tank and with another surface to the ceramic emitter.
 
The response recorded by the microphone has the typical behavior of an ideal pulse filtered by the impulse response of a physical low pass system with a bandwidth of about 16 kHz, attributed to the microphone (Fig.\ 7). In spite of the filtering, the relative energy of the pulses can be measured as a function of the angle of the normal to the diaphragm respect to the axis of propagation of the force.
	 
Relative pulse amplitude with energies averaged over four pulses per angle are shown in Fig.\ 8 and are in agreement with a possible manifestation of a vector force acting directly on the membrane. No signal has been detected outside the impact region.

\section{Discussion}
\label{dis}

\medskip \noindent
{\bf (a) Capacitance of the emitter. Current at the discharge.}

High voltage discharges like those described in Sections \ref{exp}, \ref{res} are well known in the literature. They do not require pre-ionisation, provided the electric field between the electrodes is sufficient to cause avalanche ionisation. However, the presence of a superconducting electrode causes some difference in the form and colour of the crown and of the sparkle, with respect to discharges between normal electrodes.

We recall that for the maximum voltage ($2 \cdot 10^6 \ V$) the peak value of the current is of the order of $10^4 \ A$, and the duration of the discharge varies between $10^{-5}$ and $10^{-4} \ s$. This implies a total negative charge $Q$ on the emitter, just before the discharge, up to 0.1 $C$. The associated electrostatic energy is $U = QV/2 \sim 10^5 \ J$.

If the emitter was made of a normal conductor, its capacitance could be at most $10^{-10} \ F$ (for an estimate one can consider the capacitance of a sphere, namely $C \sim 10^{-10} \cdot R$ in SI units, where $R$ is the radius). The above value for $Q$ implies a much higher effective capacitance, of the order of $10^{-7} \ F$.

During the discharge an intense super-current flows from the superconducting electrode towards the ionised gas. After the superconducting charge carriers leave the negative electrode, pairs are broken and electrons are captured by the helium ions which are striking the electrode. We thus have a peculiar kind of superconductor-normal junction. Another junction is that between the superconducting emitter and the normal metallic plate. 

From the theoretical point of view, these junctions can be described by Ginzburg-Landau models without specific reference to the microscopic theory. A typical time scale for the time-dependent Ginzburg-Landau equation in YBCO is of the order of $10^{-8} \ s$ \cite{Ummarino}, much smaller than the discharge time. This suggests that a quasi-stationary approximation could be adequate. The situation is however complicated by the following circumstances:

(1)	In anisotropic materials like HTCs the Ginzburg-Landau equation involves an order parameter with non-trivial symmetry and the pairs are thought to possess intrinsic angular momentum.

(2)	The transport current is very intense. In type-II superconductors, transport currents are associated to non-reversible displacement of flux lines, and this effect in turn depends strongly on the pinning properties of the material.
 
(3)	The emitter is subjected to a strong external magnetic field during the discharge. 

More efforts are therefore necessary, both on the experimental and theoretical side, for a complete understanding of the state of the superconducting emitter during the discharge.

\medskip \noindent
{\bf (b) Evidence of the gravitational-like nature of the effect.}

The gravitational-like nature of the effect is best demonstrated by its independence on the mass and composition of the targets (see Section \ref{res}). We are aware, of course, that gravitational interactions of this kind are absolutely unusual (see Sections \ref{bas}, \ref{kno}). For this reason, several details of the experimental apparatus were designed with the explicit purpose of reducing spurious effects like mechanical and acoustic vibrations much below the magnitude order of the observed anomalous forces. 

Indirect evidence for a gravitational effect comes from the fact that any kind of electromagnetic shielding is ineffective. Note that if one can explain in some way the anomalous generation of a gravitational field in the superconductor, its undisturbed propagation follows as a well-known property of gravity (see Section \ref{pos}). Indirect support for the gravitational hypothesis also comes from the partial similarity of this apparatus to that employed by Podkletnov for the stationary weak gravitational shielding experiment \cite{p1}.

If the effect is truly gravitational, then the acceleration of any test body on which the impulse acts should be in principle independent on the mass of the body. Suppose that $l$ is the length of a detection pendulum and $g$ is the local gravitational acceleration. Let $d$ be the half-amplitude of the oscillation. Let $t$ be the duration of the impulse, and $F$ its strength. $F$ has the dimensions of an acceleration ($m/s^2$) and can be compared with $g$. One easily computes that the product of the strength of the impulse by its duration is
\begin{equation}
	Ft = \sqrt{2gl [1-\sqrt{1-(d/l)^2} ]}
\end{equation}
If $d \ll l$, this formula can be simplified, and we have approximately $Ft \sim \sqrt{(g/l)}d=2\pi d/T$, where $T$ is the period of the pendulum. With the data of Table 1, taking $t=10^{-4} \ s$, one finds $F \sim 10^3g$.

Here, however, we encounter a conceptual difficulty. Suppose to place on the trajectory of the beam a very massive pendulum (say, $10^3 \ Kg$). If the effect is gravitational, then the acceleration of a test mass should not depend on its mass. However, it is clear that in order to give this mass the same oscillation amplitude of the small masses employed in the experiment, a huge energy amount is necessary, which cannot be provided by the device. Therefore the effect would seem to violate the equivalence principle. Considering the back-reaction is probably necessary, namely the fact that the test mass exerts a reaction on the source of the impulse. This reaction is negligible as long as we use small test masses.

\medskip \noindent
{\bf (c) Anomalous features of the observed ``radiation".}

Independently from any interpretation, the abnormal character of this radiation appears immediately clear. It appears to propagate through walls and metal plates without noticeable absorption, but this is not due to a weak coupling with matter, because the radiation acts with significant strength on the test masses free to move. Furthermore, this radiation conveys an impulse which is certainly not related to the carried energy by the usual dispersion relation $E = p/c$. A corresponding energy transfer to the test masses is not in fact observed (unless one admits perfect reflection, which seems however very unlikely).

The denomination ``radiation" is actually unsuitable, and one could possibly envisage an unknown quasi-static force field. In this way one could explain why an impulse is transmitted to the test masses. However, it is hard to understand how such a field could be so well focused.

\subsection{A possible theoretical explanation. Basic concepts.}
\label{bas}

The following Sections contain an informal introduction to a theoretical model originally developed by one of the authors \cite{m1,m2,m3,m4,m5} in order to explain the weak gravitational shielding effect by HTC superconductors \cite{p1,p2}. We suggest that this theory may be a starting point for the explanation of the impulsive gravitational-like forces described in the present paper.

The quantum properties of the gravitational field play an essential role in this model. These properties are not adequately known yet, therefore the proposed model is still in a preliminary development phase and its predictive capabilities are quite limited. Its primary merit is to envisage a new dynamic mechanism which could account for the effect. However, a full theory which justifies the model does not exist so far. This should be, in fact, a theory of the interaction between gravity (including its quantum aspects), and a particular state of matter - that of a HTC superconductor - which in turn is not completely known. 

The glossary below \footnote{{\it Cosmological term = vacuum energy density}: see (b).

{\it Lagrangian $L$ = action density}. Minimization of the action of a system gives its dynamical equations.

{\it Quantum condensate in a superconductor}: ensemble of the Cooper pairs, with collective wave function $\psi_{GL}$ (also called an ``order parameter") supposed to obey the Ginzburg-Landau (GL) equation. In the non-relativistic limit, the Lagrangian density of the condensate is just the opposite of the GL free energy density. 

{\it Zero-modes of the Einstein action = gravitational dipolar fluctuations}: see (c).

{\it Critical region}: region of the condensate with positive Lagrangian density. According to the GL theory, it can only exist for constant solutions of the GL equation or in the neighbourhood of a local relative maximum of the Cooper pair density $|\psi_{GL}|^2$. See (f).

{\it Noise source = density matrix}: a formal way to express the fact that inside a critical region the gravitational field undergoes strong dipolar fluctuations and therefore takes on at random values $h_i^2$ with probability $\xi_i$. See (h), (i).}
collects some keywords and their equivalent, also for a better orientation in the cited works.

\medskip \noindent
{\bf (a) Weakness of the standard coupling with gravity. Anomalous coupling.}

The standard coupling of matter to gravity is obtained from the Einstein equations by including the material part of the system into the energy-impulse tensor. Since the coupling constant is $G/c^4$, very large amounts of matter/energy, or at least large densities, are always necessary in order to obtain gravitational effects of some importance. This holds also at the quantum level, in weak field approximation. It is possible to quantize the gravitational field by introducing quantum fluctuations with respect to a classical background, and then calculate the graviton emission probabilities associated to transitions in atomic systems. These always turn out to be extremely small, still because of the weakness of the coupling. 

What we proved in our cited works is that a peculiar ``anomalous" coupling mechanism exists, between gravity and matter in a macroscopic quantum state. In this state matter is described by a collective wave function. Also in this state the energy-impulse of matter couples to the gravitational field in the standard way prescribed by the equivalence principle. However, the new idea is that besides this standard coupling there is another effect, due to the interference of the Lagrangian $L$ of coherent matter with the ``natural" vacuum energy term $\Lambda/8\pi G$ which is present in the Einstein equations. The two quantities have in fact the same tensorial form but possibly different sign, and it turns out that their interference can lead to a dramatic enhancement of vacuum fluctuations.

\medskip \noindent
{\bf (b) Contribution of a quantum condensate to the vacuum energy density.}

It is well known that the ``natural" vacuum energy, or cosmological term, is very small. Until recently, it was thought to be exactly null. The most recent observations give a value different from zero, but in any case very tiny \cite{m6,RPP}, of the order of 0.1 $J/m^3$. This is usually supposed to be relevant at cosmological level, in determining the curvature and expansion rate of the universe on very large scales. 

The observed value can be regarded as the residual of a complex interplay, in a still unknown high energy sector of particle physics, between positive and negative vacuum energy densities. According to \cite{Shapiro}, the observed residual should also be scale-dependent and this dependence could appear most clearly at length scales corresponding to the mass of the lightest particles like neutrinos or unidentified scalars.

In certain conditions, the action density of a Bose condensate in condensed matter can be greater than the cosmological term, but it is nonetheless very small. Its effect on the local space curvature is absolutely negligible. 

\medskip \noindent
{\bf (c) The gravitational zero modes and the polarization of the gravitational vacuum.}

Why so do we think that the interference of these two terms, which are in any case very small, can lead to some observable gravitational effect? Because we proved the existence of gravitational field configurations in vacuum, for which the value of the pure Einstein action (without the cosmological term) is exactly null, like for flat space. We called these configurations ``zero modes of the Einstein action". At the quantum level, where fluctuations are admitted with respect to flat space, they are free to grow unrestrained. In a certain sense, the quantum space-time is unstable and has a natural tendency not to stay in the flat state, but to fall into these configurations, with a definite probability. They are virtual configurations, in the sense that they are permitted in the Feynman integral describing the quantum theory. In the Feynman integral all the possible configurations are admitted - not only those satisfying the field equations - each weighed with probability equal to the exponential of its action divided by $\hbar$. All these configurations (fluctuations) take part in defining the state of the system. 

The zero-modes of the Einstein action are in all the same as fields produced by mass dipoles. In nature, mass dipoles do not exist as real sources. Nevertheless, the dipolar fluctuations mathematically have such a form. Being vacuum fluctuations, they are invariant under translations and Lorentz transformations, they are homogeneously allocated in space and at all length scales. That is, the dipolar fluctuations correspond to the fields produced by dipoles of various sizes, distributed in a uniform way. They can only show their presence if, in some way, their homogeneity and uniformity are broken. We are in presence of an analogue of the vacuum polarization in quantum electrodynamics. In that case, virtual couples of electrons/positrons pop up in the vacuum, and then quickly annihilate, generating uniform fluctuations. It is well known that these virtual processes must be taken into account, because they affect, for instance, the bare charge of the particles and their couplings. However, quantum corrections are, as a rule, very small in quantum electrodynamics. 

It is important to be aware that the gravitational dipolar fluctuations we are talking about are neither the ordinary fluctuations of perturbation theory nor the well known ``spacetime foam" fluctuations, which appear in quantum gravity at very short distances. Their nature is completely different. Although their intensity can be very large, they have null action, thanks to the compensation of positive and negative curvature between adjacent zones of space-time. The dipolar fluctuations are not zero-modes of the Lagrangian, but of the action. Their existence is possible thanks to the fact that the gravitational Lagrangian is not defined positive, unlike the Lagrangian of electromagnetic and gauge fields.

\medskip \noindent
{\bf (d) The vacuum energy cuts the gravitational zero-modes to a certain level.}

Let us go back to the cosmological term. One finds that it is related to the dipolar fluctuations, because it sets an upper limit on their amplitude. The contribution of a dipolar fluctuation to the cosmological term is typically of the form \cite{m4} 
\begin{equation}
	\Delta S = \Lambda \tau M r^2 Q
\end{equation}
where natural units are used ($\hbar=c=1$); $\tau$ is the duration of the fluctuation, $M$ is the order of magnitude of the virtual +/- masses, $r$ their distance and $Q$ is an adimensional function which depends on the detailed form of the dipole. If $\Delta S \gg 1$, then the fluctuation is suppressed.

With an electromagnetic analogy (not to be pushed too far) we could then say that the cosmological term sets the gravitational polarizability of free space. One expects that in a full non-perturbative theory of quantum gravity the bare value of the gravitational constant $G$ should be renormalized by this effect. However, as long as the cosmological term is uniform in space and time, there are no observable consequences. On the other hand, if the vacuum energy density changes locally, this can have observable effects. In particular, if there are positive local contributions which subtract from the natural density, the result can be a local increase in the gravitational polarizability.

\medskip \noindent
{\bf (e) The anomalous coupling is only active when the condensate is in certain particular states.}

Our model can also explain why only certain superconductors in certain conditions show evidence of anomalous coupling with the gravitational field. The key point is not simply the presence of a quantum condensate, nor the density of this condensate. In fact, if this was the case, anomalous gravitational effects would be observed also with low temperature superconductors, or with superfluids. But even in the static Podkletnov experiment, only in certain conditions observable effects are obtained, and their intensity is variable. 

Therefore the presence of the condensate is not sufficient to cause the effect. What are the necessary conditions? The anomalous contribution to the cosmological term is given by the Lagrangian density of the condensate, according to the following equations. Consider a scalar field $\phi$ interacting with gravity (we use units $\hbar=c=1$; SI units will be restored in eq.\ (\ref{finaleconrho2})). The interaction action is obtained from the energy-momentum tensor: 
\begin{equation}
	L = \frac{1}{2} (\partial_\alpha \phi \partial^\alpha \phi - m^2|\phi|^2)
\label{lagr} 
\end{equation}
\begin{equation}
	T_{\mu \nu} = \Pi_\mu \partial_\nu \phi - g_{\mu \nu} L =
	\partial_\mu \phi^* \partial_\nu \phi - g_{\mu \nu} L
\end{equation}
	\begin{equation}
	S_{interaction} = \frac{1}{2} \int d^4x
	 \sqrt{g(x)} T^{\mu \nu}(x) h_{\mu \nu}(x)
\end{equation}

To lowest order in $h_{\mu \nu}$ the interaction action can be rewritten as
	\begin{equation}
	S_{interaction} = \frac{1}{2} \int d^4x \left(
	h_{\mu \nu} \partial^\mu \phi^* \partial^\nu \phi
	- {\rm Tr}h  L \right)
\end{equation}
On the other hand, the cosmological term is (still to lowest order in $h_{\mu \nu}$ and
expanding $\sqrt{g} = 1 + \frac{1}{2} {\rm Tr}h + ...$)
	\begin{equation}
	S_\Lambda = \frac{\Lambda}{8\pi G} \int d^4x
	\left( 1 + \frac{1}{2} {\rm Tr}h \right)
\end{equation}
Therefore the sum of the two terms can be rewritten as
	\begin{equation}
	S_{interaction} + S_\Lambda = \frac{1}{2} \int d^4x
	h_{\mu \nu} \partial^\mu \phi^* \partial^\nu \phi
	+ \frac{1}{2} \int d^4x  {\rm Tr}h
	\left( \frac{\Lambda}{8\pi G} - L \right)
\label{sum}
\end{equation}

We see that to leading order the coupling of gravity to $\phi$ gives a typical source term $(h_{\mu \nu} \partial^\mu \phi^* \partial^\nu \phi)$ and subtracts from $\Lambda$ the local density $8\pi GL(x)$. This separation is arbitrary, but useful and reasonable if the
Lagrangian density is such to affect locally the ``natural" cosmological term and change the spectrum of gravitational vacuum fluctuations corresponding to virtual mass densities
{\it much larger than the real density of} $\phi$.

For example, suppose that $\phi$ represents a condensate with the density of
ordinary matter ($\sim 1 \ g/cm^3$). At the scale $r \sim 10^{-4}~cm$, $\tau \sim 10^{-4}~s$, with the observed value of $\Lambda$, the upper bound on the virtual source density is $\sim 10^{17} \ g/cm^3$, which is much larger than the real density. If $L$ is comparable to $\Lambda/8\pi G$ in some region, an inhomogeneity in the cut-off mechanism of the dipolar fluctuations will follow, and this effect could exceed by far the effects of the coupling $(h_{\mu \nu} \partial^\mu \phi^* \partial^\nu \phi)$ to real matter.

The value of the Lagrangian depends on the state the condensate is in, and more exactly on the macroscopic wave function $\psi_{GL}$ of the Cooper pairs. The problem of finding the wave function in given experimental conditions is still open. From the wave function one can compute the Lagrangian, and then the local contribution of the superconductor to the vacuum energy, and its ability to produce anomalous gravitational coupling. 

\medskip \noindent
{\bf (f) $L$ subtracts from $\Lambda$ only where the density $|\psi_{GL}|^2$ has a local maximum.}

The Lagrangian $L$ in eq.\ (\ref{lagr}) is the Klein-Gordon Lagrangian which describes a relativistic free scalar field. In order to relate it, in the low energy limit, to the Ginzburg-Landau (GL) free energy of superconductors \cite{m7} one makes the standard transformation
\begin{equation}
	\psi_{KG}({\bf x})= e^{imt} \phi(x)
\end{equation}
where $m$ is the Cooper pair mass. Then one turns to the GL wave function by re-normalizing $\psi_{KG}$ as follows
\begin{equation}
	\psi_{KG}({\bf x})= \sqrt{m} \psi_{GL}({\bf x})
\end{equation}
This normalization corresponds to the standard relation $|\psi_{GL}({\bf x})|^2=\rho({\bf x})$, where $\rho$ is the density of Cooper pairs. Finally, the free Lagrangian is generalized by adding a quadratic and a quartic interaction term and the minimal electromagnetic coupling. In this way one finds
\begin{equation}
	L  = - \frac{1}{2m} | -i \nabla \psi + 2e{\bf A}\psi|^2 - \alpha \psi^* \psi - \frac{1}{2} \beta (\psi^* \psi)^2
\label{gllagr}
\end{equation}
This is the Lagrangian density to be inserted into eq.\ (\ref{sum}) as contribution to the vacuum energy density. It is the opposite of the GL free energy density \cite{m8}, as expected. For wave functions which satisfy the wave equation derived from (\ref{gllagr}), the expression of $L$ simplifies to
\begin{equation}
	L = - \frac{1}{2m} \left[ \hbar^2(\nabla \rho)^2 + \hbar^2 \rho \nabla^2 \rho -m \beta \rho^2  \right]
\label{finaleconrho2}
\end{equation}
where $\hbar$ has been re-introduced to allow for a better numerical estimate.

We recall \cite{Waldram} that the $\alpha$ and $\beta$ coefficients depend on the absolute temperature $T$. The coefficient $\beta$ is always positive and approximately constant near $T_c$; $\alpha$ is negative for $T<T_c$ and near $T_c$ behaves like $const. (T-T_c)$. The ratio between $\alpha$ and $\beta$ is given by the relation $n_p=-\alpha / \beta$, where $n_p$ is the average density of pairs in the material. Finally, $\beta$ is linked to the value of the Ginzburg-Landau parameter $\kappa=\lambda/\xi$ by the relation $\kappa^2 = m^2 \beta/(2\mu_0 \hbar^2 e^2)$. Two further important points which should be taken into account are (i) the boundary conditions on $\psi_{GL}$ at the superconductor/ normal conductor and superconductor/ionized gas interfaces; (ii) the intrinsic anisotropy of HTC materials, which actually also affects the form of the GL free energy.

The relative importance in eq.\ (\ref{finaleconrho2}) of the terms with the gradients and the term proportional to $\beta$ depends on the length scale at which the variations of $\psi_{GL}$ can occur. If, as usual for Type II and HTC superconductors, this scale is of the order of $10^{-8} \ m$ or less, the terms with the gradient dominate.

It is straightforward to check that the sign of $L$ is negative, except for two types of configurations:

(1)	For the constant solutions of the Ginzburg-Landau equation in the absence of external field, namely $\rho({\bf x})=n_p$. The corresponding constant Lagrangian density is $L_1=\frac{1}{2} \beta n_p^2$.

(2)	For regions of the condensate where $\rho \nabla^2 \rho$ is negative and greater, in absolute value, than $(\nabla \rho)^2$. It is straightforward to check that these are regions located around local density maximums, or more generally about lines and surfaces where the first partial derivatives of $\rho$ are zero and the second derivatives are negative or null. The Lagrangian density at a maximum is $L_2 \sim \frac{1}{2m} \rho |\rho ''|$. If the maximum is sharp, $L$ can be much larger than for constant solutions. Configurations of this kind are characteristic of solutions of the Ginzburg-Landau equation with strong magnetic flux penetration \cite{Tilley}.

Now we make some numerical estimates of $L$. Assuming for YBCO $\rho \sim 10^{27} m^{-3}$, $\xi \sim 10^{-9}~m$, $\kappa \sim 10^2$, we obtain for the configurations (1) and (2) densities $L_1\sim 10^4 J/m^3$ and $L_2\sim 10^7 J/m^3$, respectively. With reference to our experiment we must explain, on the basis of these figures, why anomalous coupling at the emitter takes place, but only at the moment of the discharge and in presence of strong magnetic flux penetration.

We remind that the vacuum energy density acts as cut-off for the dipolars fluctuations, so that their maximum amplitude $A $ is inversely proportional to $ \Lambda $. Therefore in presence of a local contribution by the superconductor we have
\begin{equation}
A \propto \frac{1}{|\Lambda/8\pi G-L|}
\end{equation}
So an amplification of the fluctuations is possible when $L $ has the same sign as $ \Lambda $. From the experimental observations it is clear that a negative value of $L $ does not give the necessary conditions. In fact, in that case anomalous coupling would be observed every time there are strong density gradients in the superconducting material - which occurs almost always. From this we obtain a first piece of information about the value of $ \Lambda $ at atomic scale (otherwise unknown): it must be positive. $ \Lambda/8 \pi G $ must furthermore be much larger than $L_1 $, because a constant density does not give observable anomalous effects either. In conclusion, $ \Lambda/8 \pi G $ must be larger than $L_2 $, and the amplitude $A $ of the dipolar fluctuations increases as the local value of $L $ approaches it from below. 
\footnote{The density $L_2$ can be compared with that obtainable for an electromagnetic field alone - which can also be regarded as ``coherent matter" in this context. The electromagnetic energy density is proportional to $({\bf E}^2+{\bf B}^2)$, but the action density is proportional to $({\bf E}^2-{\bf B}^2)$, therefore only an electrostatic field has the right sign. But even with a field strength of $3 \cdot 10^6 \ V/m$, one obtains a density of 100 $J/m^3$, definitely lower.}

This can happen in presence of strong magnetic flux penetration (Point 2 above), but that requisite is not sufficient: the presence of a strong transport current is also necessary. In fact, with strong flux penetration, the density $ \rho $ is depressed. The current appears like a means to obtain at once a large $ \rho $ and a large value of $ \rho ''$ in correspondence of the maximums. Indeed, in inhomogeneous materials like HTCs, a very intense transport current inevitably forces a large density in certain regions. In the case of the weak gravitational shielding \cite{p1} the transport current is mostly due to the accelerated rotation of the disk \cite{London}. In the present case it is due to the discharge.

\medskip \noindent
{\bf (g) The anomalous coupling inside the condensate modifies the field also outside.}

Summarizing, we can say that in particular conditions a quantum condensate is able to modify locally the cosmological term of the gravitational action. In turn, the cosmological term fixes the cut level of the dipolar fluctuations, and so the local gravitational vacuum polarizability. Therefore an anomalous coupling of the condensate with the gravitational field is observed. This coupling can be abnormally strong, due to the large amplitude of the vacuum fluctuations. It is a kind of coupling that completely eludes the standard form and so is not proportional, through the $G/c^4$ constant, to the energy-impulse content of the condensate. 

The next question is: how is the gravitational field modified inside the condensate and in its neighbourhood? In other words, let us suppose that the gravitational field displays a much higher level of dipolar fluctuations in some regions of the condensate. What is the consequence for the global behaviour of the field? For instance, with reference to the static experiment, why is there a noticeable reduction of the terrestrial field above the disc? We can say that in that case there is a big gravitational source - the earth - producing a uniform field in the region of interest, according to the usual classical field equations. Into these equations, however, we must also introduce a kind of ``random source" located in the condensate. This affects the field also outside. 

\medskip \noindent
{\bf (h) The modified field equation contains an arbitrary constant.}

We have proposed to introduce into the gravitational field equations a term representing a random source located in the condensate. This procedure has some intrinsic arbitrariness. The intensity of the source is arbitrary, too, and cannot be obtained from the ratio between the energy density of the condensate and the ``natural" vacuum energy density. In fact, it is true that this ratio allows us to determine the maximum amplitude of the dipolar fluctuations. However, the probability of these fluctuations, and so their weight with respect to the external configurations, is still an unknown element of the theory. We know that certain field configurations give a null contribution to the pure Einstein action and can grow up without limit, and we can determine the cut-off amplitude due to the intervention of the cosmological term; however, the probability that these fluctuations really happen is a feature of the quantum theory, which is not known yet. In principle, it would be possible to write a functional integral containing all the involved fields, the test masses etc., evaluate it and obtain everything. In practice, what we can do with such an integral is only a weak field expansion about the minimum of the action.

In conclusion, if we represent the gravitational fluctuations in the quantum condensate as a random source, the intensity of this source must be inserted as a parameter which is fixed ``a posteriori" from the experimental data. This should not be unexpected. The experimental observations described in this work are absolutely new and unprecedented. They inform us about a sector of the theory which is otherwise unknown. The validity of our model consists of suggesting a new interaction process, while avoiding contradictions with known facts, and while giving the freedom to fix a new coupling constant. 

\medskip \noindent
{\bf (i) The modified field equation must be able to describe the observed phenomenology.}

After modelling the effect of the fluctuations on the external field (for instance, as random source with a certain strength), we are able to verify the consistency and correctness of the procedure by calculating this field and comparing the result to the data - not only as intensity but also as shape. For instance, we know that in the static experiment a ``shielding cylinder" was observed which extended far above the superconducting disc and showed net borders, without diffraction. Obtaining this characteristic from the field equations modified by the anomalous coupling is all but trivial. The observed field configuration is furthermore clearly not conservative. (This can be understood remembering that there is a source term in the equation.) Nonetheless, we succeeded in \cite{m3} in obtaining a result of this kind (see also \cite{nat}).

\medskip \noindent
{\bf (j) Euclidean and Lorentzian formulation of the anomalous coupling and of the modified action.}

From the technical point of view, the general program outlined above has been implemented in the Euclidean (or imaginary-time) version of quantum gravity \cite{m1,m2,m3}, and more recently in the standard Lorentzian version \cite{m4}.

The occurrence of the dipolar fluctuations and the cut-off role of the vacuum energy density are most clearly and safely exhibited in the Lorentzian formalism. On the other hand, the Euclidean formulation is reliable and more suitable for perturbative calculations of the modified field equations.

The Lorentzian formulation is indispensable for the latest developments (calculation of the low-energy limit of $L$ and determination of its sign as a function of $\rho$).

\subsection{Possible interpretation of the gravitational-like impulse at the discharge}
\label{pos}

We focus now our attention on the new, puzzling effects described in the experimental part of this paper, and partially analysed at the beginning of this Section. We cannot offer any quantitative interpretation of these effects yet, so we shall limit ourselves to an hypothesis based on our previous work on the static effect. 

\medskip \noindent
{\bf (a) From the static to the transient effect: the virtual dipoles emit virtual radiation.}

We have seen that the observed impulse can neither be described as a real radiation, nor as a quasi-static field. These are the two types of field configurations which are predicted by any classical theory.

However, we also saw that the weak gravitational shielding effect can be explained introducing the concept of anomalous dipolar quantum fluctuations induced by the condensate. These fluctuations locally increase to observable levels the gravitational vacuum polarizability.

In the transient case, the virtual dipolar fields are rapidly varying in time, because the critical conditions in the quantum condensate are produced only for a short time at the discharge. Therefore it is natural to expect that in this case the formation of virtual dipoles is accompanied by the emission of virtual dipolar radiation.

\medskip \noindent
{\bf (b) In quantum mechanics every interaction is equivalent to an exchange of virtual radiation.}

Let us recall the concept of ``virtual radiation" as intermediate state of a quantum process. In quantum mechanics, every interaction is thought to take place through the exchange of virtual particles - photons, gauge bosons, gravitons. This is especially clear in scattering processes, but also the static interaction potential can be written as an integral over time and four-momentum of the propagator of a mediating virtual particle \cite{m9}. In this integral, all values of energy-momentum are included, not only the ``on-shell" values satisfying the condition $E=p/c$ which is typical of real photons and gravitons. This is possible just because the mediating particles are virtual, i.e.\ so short-living that the Heisenberg principle allows for a large indetermination in their energy. In fact, the static force obtained as a sum of virtual processes is very different from the effect of a collection of free photons or gravitons. 

It is also possible to compute along these lines the quantum corrections to the classical force. To this end, one includes in the sum some less probable processes, involving the exchange of several mediating particles at the same time, or the creation-annihilation of virtual pairs, and so on.

\medskip \noindent
{\bf (c) The impulse at the discharge is made of off-shell gravitons.}

In order to explain the weak gravitational shielding effect we hypothesised an anomalous strong quantum correction to the field of the earth. In the transient case, on the other hand, all the emitted ``radiation" should be regarded as a quantum effect. The observed impulsive force could be the effect of virtual processes in which:

1.	The intense supercurrent flowing across the superconducting cathode produces critical gravitational regions, with strong dipolar vacuum fluctuations. The necessary conditions are those discussed in Section (\ref{bas}-f).

2.	These fluctuations emit a beam of virtual off-shell gravitons with $E \ll p/c$. The direction of the beam is sharply defined by momentum conservation. Whatever the exact microscopic mechanism, the momentum carried by the virtual radiation can only originate from that of the Cooper pairs entering the critical regions, i.e.\ from the momentum of the supercurrent. This is in agreement with the fact that if the impact direction of the discharge changes (like in the earlier set-up, see \ref{res}), also the direction of the emitted radiation changes.

3.	When the beam hits a mobile target, it  conveys an impulse to this target. Like for usual gravitational forces (which can also be interpretated as graviton exchange, as mentioned), interposed bodies do not affect transmission. 

\subsection{Known effects which could be connected to the observed phenomenon.}
\label{kno}

The following concepts are often quoted to explain anomalous gravitational effects. So we summarise them quickly, even though in our case they do not fit to the observed phenomenology.

We recall that several alternative theories were proposed, in the last decades, in the attempt to ``explain" and give a physical basis to spacetime and vacuum. Every one of these, to be credible, must first of all reproduce the known results of general relativity and quantum field theory. Notable examples are theories of ``induced" gravity according to Sakharov's idea \cite{Barcelo}, and string theory. It is not easy to judge if anyone of these theories can also predict, in addition, unusual effects in the presence of superconductors. Some are not developed enough to give a definite answer. Many have been proposed recently, still their authors did not envisage any anomalous coupling to matter in a macroscopic quantum state. In general, if the coupling of gravity to matter conforms to the classical Einstein equations, then the intensity of any generated gravitational field will depend only on the energy-momentum of the source multiplied by $G/c^4$, so there is no chance for anomalous effects. A notable exception could be torsion theory, which predicts a coupling to quantum spin not allowed in the Einstein equations; however, as we shall see below, there are strict experimental limits on the coupling of matter to torsion. Another possibility is to consider strong quantum effects, as done in Section \ref{pos}; in that case, we chose to start from the standard form of the classical theory.

\medskip \noindent
{\bf (a) The gravitomagnetic field}

Could the observed anomalous forces be due to the fact that the superconductor produces during the discharge a strong gravitomagnetic field? It is well known from general relativity that the gravitational field contains components of magnetic type, called the gravitomagnetic field. These components have the property to be produced by moving objects and to act on objects in movement. It is possible to write the Einstein field equations in weak field approximation in a form very similar to that of Maxwell equations. The important difference, with respect to the Maxwell equations for electromagnetism, is that gravitomagnetic effects are suppressed by factors $1/c$ or $1/c^2$. So they are always very small with respect to ``gravitoelectric" effects - those which have a Newtonian limit. See for instance \cite{Paik,Ciufolini}; in these references an analysis of the Maxwell equations for gravity and also of the ``Gravity Probe B gyroscope experiment" for detecting the Earth's gravitomagnetic field is given.

In certain conditions the gravitomagnetic field can be repulsive. In neutrons stars it can produce a gravitational analogue of the Meissner effect. As one would expect, however, the ``gravitational Meissner effect" is exceedingly weak: for instance, it has been computed that in a neutron star with a density of the order of $10^{14}\ g/cm^3$, the London penetration depth is ca. 12 km \cite{Lano}. In the 1970s experiments were done to detect possible weight decreases of a rotor while it was rotating at very high speed \cite{Hayasaka}, but conclusive results were never obtained. 

\medskip \noindent
{\bf (b) Gravitomagnetism and quantum spin}

The aspects of the theory of quantum gravity concerning the spin are summarized, for instance, in \cite{Barker}. In previous work \cite{Oconnell} O'Connell examined the implications arising from the fact that spin contributions to the gravitational potential are as large as the spin-independent contributions for interparticle separations of the order of Compton length and also that such spin-dependent forces could be repulsive for certain spin orientations. An analysis was also carried out of the possibility of measuring gravitational spin-dependent forces in the laboratory \cite{Rasband}. An idea of one such experiment, as proposed in \cite{Peres}, is to observe a breaking of the equilibrium of a polarized body, hanging in the gravitational field, when its polarized state is destroyed. This effect, if present, is predicted to be extremely small. (See also the results by Ritter et al.\ \cite{Ritter}.) Claims like those by Wallace of ``antigravity" forces produced by spin orientation in nuclei \cite{Wallace} appear as totally unreliable today.

\medskip \noindent
{\bf (c) The works by Ning Li and Torr}

In a series of articles Ning Li and Torr \cite{p9} calculated the gravitomagnetic field which would be produced by a superconductor containing circulating supercorrents. According to them, in this case also the movement of the ions, which produce a current of mass, has particular importance. Also the fact that spin alignment is present would be very important. The alignment of the spin of the lattice ions would be a source of gravitomagnetic field. The objections to this model are of two types. First, the total spin amount that is possible to obtain in normal condensed matter is always very small, also in presence of alignment. Therefore, given the weak coupling with the gravitomagnetic field, one does not understand how it could reach detectable intensity. Another objection is of more technical nature. Ning Li and Torr use the Maxwell equations for gravity, which hold in weak field approximation. However, when they find that some terms of these equations ``explode", they keep this result even if it is inconsistent with the initial approximation. 

Previously to the first Podkletnov experiment \cite{p1} and without reference to it, Ning Li and Torr determined another consequence of their model \cite{p10}. Not only the alignment of the spin of the lattice ions would produce a gravitomagnetic field, but in the presence of an outside applied time-dependent vector potential, this would be converted into a perceptible gravitoelectric field. Such a conversion is necessary, if one wants to apply this theory to the weak gravitational shielding, because in that case the test bodies are at rest, and so they would not respond to a gravitomagnetic field. However, this is in contrast with the short-lived permanence of the effect also after all the outside fields are turned off. At last, we notice that Ning Li and Torr never published a work in which they try to interpret the phenomenology of the effect in terms of their model nor to explain why it is observed only with certain types of superconductors.

\medskip \noindent
{\bf (d) Other models connected to gravitomagnetism}

There have been proposals of alternative interpretations of gravitomagnetism in connection with the Mach principle \cite{Nordtvedt}. In this context, the gravitomagnetic phenomena would achieve a greater importance. Still with reference to a Machian theory, Woodard claims to have experimentally obtained some transient fluctuations in the inertial mass of a capacitor \cite{Woodward}. However we will not be concerned here with these approaches, which to a large extent remained at a subjective level.

Recently Ummarino \cite{Ummarino} tried to establish a link between the Podkletnov effect and the gravitomagnetic and gravitoelectric field, following a more standard approach, connected with the Ginzburg-Landau theory for superconductors.

\medskip \noindent
{\bf (e) Theories with torsion}

We now come to the theories of gravity with torsion. Could the observed phenomenology, not understandable in general relativity, be related to the existence of a torsion field? Let us first recall how the concept of torsion was born: from the idea of considering the connection (which in the metric formalism is simply a quantity derived from the metric and is symmetric with respect to a couple of indexes) like an independent quantity, with its own dynamics and, generally, not symmetric. Torsion theory is an extension of general relativity, which was investigated in great detail in the past. It is necessary in fact in order to introduce the interaction of the gravitational field with the quantum spin, and allows to connect general relativity to the usual gauge theories. After admitting the possibility of the existence of torsion, its features (for instance, its propagation), do not descend from first principles or from the geometrical structure of the theory, but from the form of the various possible terms in the action and from coupling constants fixed on the basis of experimental observations. Over the years a large amount of experimental data have been accumulated, which place strong limits on the couplings.

A picture of the current situation is for instance presented by Carroll and Field \cite{Carroll}. What is concluded is that the possible existence of the torsion can be of interest at the level of gravitational interactions at very short distances (Planck scale), but not at the level of laboratory experiments. 

In their work Carroll and Field discuss possible actions for torsion and its interaction with matter fields like those of the standard model of particle physics. They construct a free Lagrangian from powers and derivatives of the torsion, and couple it ``minimally" to matter through the covariant derivative. They find that there is only a small range of models possible without placing arbitrary restrictions on the dynamics. In these models only a single mode interacts with matter, either a massive scalar or a massive spin-1 field, and each mode is parameterised by two constants with the dimension of mass. They concentrate on the scalar theory, which is related to several proposals found in the literature and discuss what regions of parameter space are excluded by laboratory and astrophysical data. Carroll and Field find that a reasonable expectation  would be for each of the two mass parameters to be of the order of the Planck scale; such a choice is a safe distance away from the regions excluded by experiment. They conclude that, while there are reasons to expect that the torsion degrees of freedom exist as propagating fields, there is no reason to expect any observable signature from torsion.

Other ``not orthodox" points of view are represented in literature \cite{Akimov}. But even though theories of torsion exist since a long time, no attempts to explain the phenomenology of the Podkletnov effect were ever presented.

\medskip \noindent
{\bf (f) The value of $G$ and the measurements at short distances}

The fundamental characteristics of the gravitational interaction keep on being a very alive and interesting research field. We do not mean by this the consequences of gravity on the structure of the universe, and so the applications of general relativity to black holes, astrophysics etc.; we mean the basic features of the force, including the issue whether general relativity is an adequate and complete description of it. We recall that only two predictions of general relativity have not been verified yet, namely the existence of gravitational waves and of gravitomagnetic fields. There are furthermore several current experiments to detect possible violations of the equivalence principle. Up to now no contradiction has ever been observed with respect to the predictions of general relativity \cite{RPP}. 

The precision with which the value of $G$ is known is clearly unsatisfactory compared to the precision with which the other fundamental physical constants are known. This is also due to the fact that while the definition of the other fundamental constants relies on microscopic experiments performed with high precision devices, for the $G$ constant it is necessary to use more or less sophisticated versions of the Cavendish experiment. In particular, we do not know the behaviour of gravity at short distances (millimeters or less). According to Gillies \cite{Gillies1}, in the second half of 1900 more measurements of $G$ were made than ever before. Some discrepancies were observed in recent times with respect to the best official value fixed in 1982.

The value of $G$ has been called into question by new measurements from respected research teams in Germany, New Zealand, and Russia \cite{Gdata}. The new values disagree wildly. For example, a team from the German Institute of Standards obtained a value for $G$ that is 0.6\% larger than the accepted value; a group from the University of Wuppertal in Germany found a value that is 0.06\% lower and one at the Measurement Standards Laboratory of New Zealand measured a value that is 0.1\% lower. The Russian group found a space and time variation of $G$ of up to 0.7\%. The collection of these new results suggests that the uncertainty in $G$ could be much larger than originally thought. This controversy has spurred several efforts to make a more reliable measurement of $G$.

A recent theoretical prediction suggests that gravity penetrates extra, compact dimensions so that the gravitational inverse square law must be modified at short ranges (less than 1 $mm$). Arkani-Hamed, Dimopoulos, and Dvali \cite{Dvali}, have offered this as an explanation of the hierarchy problem. A team at the university of Washington is doing new measurements in order to check this, but results are negative up to now \cite{Adelberger}.

\medskip \noindent
{\bf (g) The anomalous acceleration of the Pioneer and scalar-tensor theories}

Let us now pass from the shortest distances at which it is possible to observe gravitational interactions in the laboratory, to the largest. Surprising results were obtained recently by the observation of the motion of the Pioneer space probes \cite{Pioneer}. Direct measurements are possible thanks to the radio signals transmitted by the spacecrafts, which give precise information about their position, speed and acceleration. From the analysis of the data a residual acceleration was found, not explainable through the usual orbit tracking models. Some proposals for a theoretical explanation of the phenomenon were presented. One of these \cite{Mbelek} calls for the existence, in addition to Einstein's gravitational field, of a scalar field which, supposed it has certain properties, could lead to the observed effect. This field could also be responsible for the rotation anomalies observed in the galaxies, because it gives a modification of the behaviour of gravity at large distances. The cited model belongs in practice to a wide group of extensions of the standard theory of gravitation \cite{Steinhardt}, which originate from the Brans-Dicke model and in more recent times refer to string theory and to the so-called dilaton field.

\medskip \noindent
{\bf (h) Gravitational anomalies at solar eclipses}

There is a long history of claims of possible anomalous changes of gravity during sun eclipses. These would be different from tidal changes, which are well known and calculable with precision, and could show that some kind of shielding of the solar attraction by the moon exists, at the moment when it steps in between the earth and the sun. The data, however, are quite contradictory \cite{Eclipse}.

The original experiments by Allais, in which the change of the oscillation period of a pendulum was observed, seemed to imply a decrease in $g$ of about 1/100 of the solar gravity. The consequence would be an apparent increase of the terrestrial attraction of about one part in a million. During the eclipse of 2000 a net of observers with pendulums and gravimeters was organised by initiative of the NASA, in order to observe possible changes; the definitive results are not available yet. A work was published in 2000 by a Chinese team, with reference to the eclipse of 1997 \cite{EclipseChina}, in which data measured with a high precision gravimeter are reported. Also in this case, some anomalies were observed at the eclipse, but with features very different from those reported before. There is a noticeable decrease of the terrestrial attraction, that is, an apparent anti-screening of the solar gravity. The shielding factor is much smaller than the one declared by Allais, and the time sequence rather strange, with maxima at the beginning and the end of the darkness period. 

It is important to remember that other experiments clearly exclude any shielding of gravitation of the so-called ``Majorana" type (these phenomena were first investigate by Majorana at the beginning of 1900) \cite{Gillies2}. Besides, one would expect a possible shielding effect by the moon to have an analogue for artificial satellites, namely a shielding of solar gravity when the satellite enters the shadow of the earth. Anomalies of this type were actually reported \cite{VanFlandern}, though they are very difficult to confirm, because the irregularities which affects the motion of the artificial satellites are numerous, especially when they are close to the surface of the earth.

\medskip \noindent
{\bf (i) Weak gravitational shielding by superconductors}

In this variegated field of investigations on possible gravitational anomalies, the work by E. Podkletnov stands out, originally appeared in 1992 \cite{p1}, then in improved version in 1995 \cite{p2}. It describes possible gravitational anomalies caused by HTC superconductors. The observed anomalies reached a maximum of about 2\% of $g$, in transient situations, and 0.3-0.5\% of $g$ in almost stationary form. These anomalies were produced by means of discs with multiple layers, rotating at high speed, in very particular conditions, which still have not been completely duplicated. A NASA team already produced a first simplified reproduction of the experiment in 1997 \cite{p3}, and a new version should start giving some data this year. Besides the remarkable experimental difficulties, the results are puzzling under the theoretical point of view. In fact, the strength of the anomalies is very large with respect to what has been previously observed, and no sufficiently complete theory exists, which can explain this kind of phenomena (see Section \ref{bas}). Here in fact both gravitation is called into play, of which a complete quantum theory does not exist, and typically quantum phenomena (behaviour of the macroscopic wave functions in superconductors). Furthermore, not conventional superconductors are involved, but HTC superconductors, for which several issues are still unsettled, like the pairing mechanism etc. 

\medskip \noindent
{\bf (j) Terahertz radiation emission by HTC superconductors}

A new type of terahertz radiation was discovered by Japanese scientists by 
irradiating HTC superconductive films with femtosecond laser pulses \cite{jap1,jap2}. The 
radiation mechanism is thought to be connected with the ultrafast 
supercurrent modulation by the laser pulses, which induces nonequilibrium 
superconductivity. The principal design of the experimental installation has 
some common features with our gravity impulse generator and the behavior of the 
superconducting crystallized materials might have a similar origin.

\section{Conclusions}
\label{con}

The experimental apparatus has shown that an impulse of gravitational-like force freely propagating through different physical media can be generated by a dual layered YBCO HTSC under pulsed electric current. The impulse propagates parallel to the direction of the discharge and orthogonal to the surface of the HTSC. The intensity of the impulse has been found to increase with increasing discharge energy, and to depend on the chemical composition and structure of the HTSC and on its internal magnetic state. In a typical measurement, the mechanical energy imparted by the impulse to a pendulum of mass $18.5~g$ was between $4 \cdot 10^{-4}~J$ and $23 \cdot 10^{-4}~J$ (Table 1).

From the theoretical point of view, we understand the effect as the result of an anomalous interaction between a special class of gravitational vacuum fluctuations and the macroscopic wave function of the superconductor. This interaction is locally activated when the product of the pairs density $\rho$ and its second derivative $\rho ''$ is sufficiently large. The sign of $\rho ''$ must furthermore be positive. These conditions can be satisfied in the presence of intense transport current and magnetic flux penetration.

Attempts of scientists to control gravity have been present for a long time and in the future these efforts will almost certainly become more prominent. Albert Einstein spent the last years of his life trying to integrate gravity with the other laws of physics and the entire scientific community remains intrigued with the problem of gravity ever since. However, since our knowledge of gravity is poor in comparison to that of the other fundamental forces, we are unable to control it in any fundamental way. We are therefore left with the option of carrying out experiments based on new theories, on scientific intuition and careful analysis of previous results. This work indicates that  a kind of artificial gravity can be generated using the unique properties of superconducting ceramic materials and a combination of electric and magnetic forces.

\bigskip \noindent {\bf Acknowledgment} - G.M.\ was
 supported in part by the California Institute for Physics
 and Astrophysics via grant CIPA-MG7099.

\newpage

\noindent
{\bf Figure Captions}

\noindent 
Fig.1	Initial setup of the impulse gravity generator.

\noindent 
Fig.2	Improved variant of the impulse gravity generator.

\noindent 
Fig.3	Discharge chamber of the impulse gravity generator.

\noindent 
Fig.4	Arkadjev-Marx high-voltage pulse generator.

\noindent 
Fig.5	Pendulum in a glass cylinder under vacuum. (The actual cylinder is wide enough to allow a complete oscillation.)

\noindent 
Fig.6	Correlation between the voltage discharge and the 
deflection of the pendulum.

\noindent 
Fig.7	Impulse recorded by the microphone at a 67 deg. impact angle. Time scale is sampling periods at $f_s=44.1~kHz$. There is a 50 $Hz$ noise due to power grid. The signal is unfiltered.

\noindent 
Fig.8.	Relative pulse impulse versus impact angle.
		
\end{document}